# Multi-User MIMO with flexible numerology for 5G

Sridhar Rajagopal and Md. Saifur Rahman
Ranzure Networks and Samsung Research America - Dallas
Dallas, TX, USA
{sridhar.rajagopal@ranzure.com, md.rahman@samsung.com}

*Abstract*— Flexible numerologies are being considered as part of designs for 5G systems to support vertical services with diverse requirements such as enhanced mobile broadband, ultra-reliable low-latency communications, and massive machine type communication. Different vertical services can be multiplexed in either frequency domain, time domain, or both. In this paper, we investigate the use of spatial multiplexing of services using MU-MIMO where the numerologies for different users may be different. The users are grouped according to the chosen numerology and a separate pre-coder and FFT size is used per numerology at the transmitter. The pre-coded signals for the multiple numerologies are added in the time domain before transmission. We analyze the performance gains of this approach using capacity analysis and link level simulations using conjugate beamforming and signal-to-leakage noise ratio maximization techniques. We show that the MU interference between users with different numerologies can be suppressed efficiently with reasonable number of antennas at the base-station. This feature enables MU-MIMO techniques to be applied for 5G across different numerologies.

*Keywords—-5G, numerology, MU MIMO*

## I. INTRODUCTION

As part of the 5G evolution of communication technology, 3GPP is developing a "New Radio (NR)" access technology which operates in frequency ranging up to 100 GHz and requires supporting a wide range of 5G use cases such as enhanced mobile broadband (eMBB), ultra-reliable low-latency communications (URLLC), massive machine type communication (mMTC) in a single technical framework [1].

These various use cases for 5G new radio interface have diverse requirements in terms of data rates, latency, and coverage. eMBB is expected to support peak data rate (20Gbps for downlink and 10Gbps for uplink) and user experienced data rates in the order of three times IMT-Advanced. On the other hand, in case of URLLC, the tighter requirements are put on ultra-low latency (0.5ms for UL and DL each for user plane latency) and high reliability ($10^{-5}$ within 1 ms). Finally, mMTC requires high connection density (1,000,000 devices/km$^2$ in urban environment), large coverage in harsh environments (164 dB Mutual Coupling Loss), and extremely long-life battery for low cost devices (15 years) [2]. To support these use cases with diverse requirements in a single technical framework, the system design needs to take into account various aspects, such as waveform and multiple access scheme, numerology and frame structure, and time-frequency resource allocation. Numerology considerations here include subcarrier spacing, symbol length, FFT size, TTI, etc. In order to support multiple numerologies, different vertical services can be multiplexed in either frequency domain, time domain, or both, for example, see [3]. Furthermore, in addition to traditional time and frequency multiplexing, spatial domain multiplexing between different vertical services could be another alternative worth investigating, given the possibility of enhanced spatial separation in 5G new air interface with large number of antennas at the eNB.

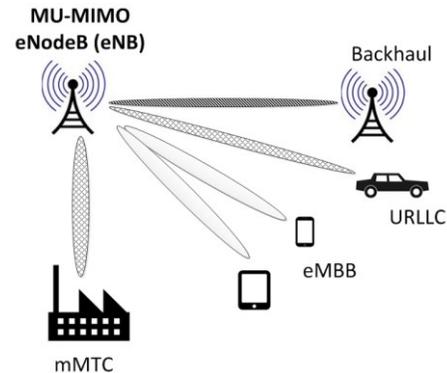

Fig. 1. 5G applications requiring support for flexible numerology

Spatial multiplexing using MU-MIMO is becoming important as a key enabler to increase capacity for 5G [4, 5]. One of the limitations to use this technology for 5G is that (MU-) MIMO operation by default assumes the same numerology (sub-carrier spacing, TTI etc.) for all streams. This paper provides insights on how to extend MU-MIMO for supporting flexible numerology. Fig. 1 shows the applications enabled by extending MU-MIMO to support users with diverse requirements such as in-band backhaul, vehicular communication, enhanced mobile speeds, mission critical services and industrial automation using the same time and frequency resource to improve capacity and providing low latency support. The eNB can communicate with various user types such as another eNB, a car and a mobile device, each with different numerology but using the same time and frequency resources. For example, the eNB can communicate with another eNB using a numerology optimized to have a larger CP length and larger subcarrier spacing in order to support a longer distance and a large bandwidth. The eNB may communicate with cars using a numerology, where the subcarrier spacing and CP length may be optimized for high speed and smaller bandwidths. The eNB may communicate with mobile devices using the regular CP and subcarrier spacing.

In particular, this paper makes the following contributions:

(1) Extending MU-MIMO design to support multiple numerologies
(2) Pre-coder design for MU-MIMO with multiple numerologies
(3) Capacity analysis to demonstrate potential gains
(4) Performance evaluations to demonstrate benefits assuming ideal channel knowledge at the transmitter

## II. PRECODER DESIGN

We provide a technique for supporting flexible numerology for UE sub-carrier spacing, CP length and bandwidth in multi-user MIMO systems, where

(a) the users are grouped according to the chosen numerology,
(b) a different pre-coder and FFT size is used per numerology at the transmitter, and
(c) the pre-coded signals for the multiple numerologies are added in the time domain before transmission.

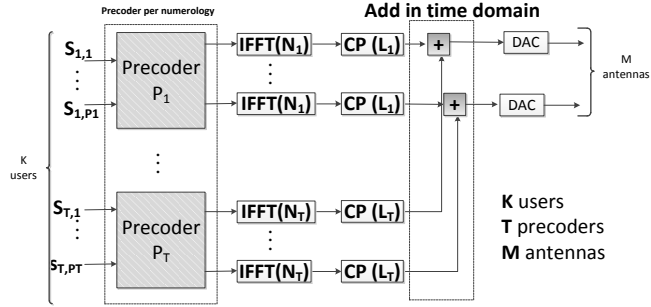

Fig. 2. MU pre-coder design with support for flexible numerology

Fig. 2 shows a MU-MIMO configuration supporting multiple numerologies. In this configuration, $K$ users are present, each of which is served with a single stream (i.e., a rank-1 transmission). The $K$ users are grouped into $T$ groups, each of which is configured with a different sub-carrier spacing and/or CP length. Let the users be denoted as $s_{t,u}$ where $t$ is the group number and $u$ is the index of the user within the group. Each group of users is precoded separately in the frequency domain using a precoder $\boldsymbol{P}_t$, which is matrix of dimension $M \times K_t$, where $K_t$ is the number of users in user group $t$. Each precoder output is converted to the time-domain using an IFFT size of $N_t$ depending on the desired sub-carrier spacing and a cyclic prefix (CP) is added of the desired length $L_u$. The output of all the different groups is then added in the time-domain and then sent to the RF with $M$ antennas for transmission. It is assumed that the users can be spatially separated for MU-MIMO operation based on eNB scheduling and grouping of users. The receiver does not have any knowledge of the interference or existence of other users, i.e., similar to MU-MIMO operation in LTE, the proposed MU-MIMO transmission scheme is transparent to the users. Hence, no special enhancements are considered at the receiver.

The transmission of the user's data is aligned to the symbol of the users with the smallest sub-carrier spacing (largest symbol length) as shown in Fig. 3. In the figure, a user group with the smallest sub-carrier spacing is denoted as $s_{1,x}$, and a user group with second smallest subcarrier spacing is denoted as $s_{2,x}$, and so on. The length of an OFDM symbol of a user group 1 is configured to be the same as an integer $p_t$ multiple of the length of an OFDM symbol of a user group $t$: $p_t(N_t + L_t) = N_1 + L_1$. The longest OFDM symbol length, i.e., $N_1 + L_1$, is denoted as a time unit. Furthermore, the starting times of the first OFDM symbol of the OFDM symbols belonging to different user groups are aligned (synchronized) for all $t = 1, \ldots, T$. This simplifies the frame structure design to allow multiple UEs with variable sub-carrier spacing and CP lengths to be allocated in a frame.

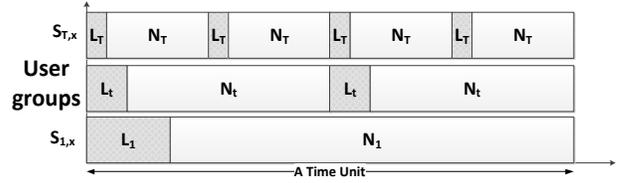

Fig. 3. Alignment of different sub-carrier spacings for MU-MIMO

In addition, it may also be beneficial to have the sub-carrier spacings of the user groups be multiples of each other to help make FFT sizes easily configurable in hardware and to support re-sampling (interpolation or decimation) of channel estimates in the frequency domain to adapt to the variable sub-carrier spacing, i.e. $p_t \cdot N_t = N_1$. Note that it is not necessary for the CP lengths to be reduced with sub-carrier spacing. The CP length can also be fixed to that of the largest CP requirement in the cell, i.e. $L_t = L_1$ for all $t = 1, \ldots, T$, where $L_1$ is decided based on the delay spread requirement to support all UEs in the cell. However, as expected, the CP overhead for the low latency/larger sub-carrier spacing users will be higher in this case.

Due to the use of different sub-carrier spacings, the channel estimate for a particular UE in the frequency domain received at the eNB or base-station will have different resolution, depending on the sub-carrier spacing group of the UE. This variation in the estimate of the channel estimation resolution needs to be accounted for in the design of the pre-coder for the different user groups. The channel estimates $h_1$ and $h_2$ for 2 users in the same bandwidth can have different frequency resolutions, depending on the sub-carrier spacing. For a UE with shorter sub-carrier spacing, the channel estimate in the frequency domain will have higher resolution compared to the wider sub-carrier spacing UE. Furthermore, the frequency locations of the first (lowest-indexed) subcarrier of the $T$ user groups are aligned (synchronized) for all $t = 1, \ldots, T$.

**Conjugate beamforming precoding**

In case of conjugate beamforming (CB), each user's estimate of the channel is conjugated and used as the pre-coder for that user and all other users are ignored, irrespective of their numerology choice. Hence, the CB pre-coder is not impacted by this design. So,

$$P_j = h_j^H, \quad (1)$$

where $P_j$ is the precoder for the channel estimate $h_j$, where $j$ is the sub-carrier index. Since the channel estimate is already at the correct resolution for the desired sub-carrier spacing for the user, there is no additional interpolation or decimation of the estimate needed in this case.

**Proposed SLNR precoding**

More advanced precoding schemes such as precoding using Signal to Leakage Noise Ratio Maximization (SLNR) [4, 5] or MMSE based techniques use the channel estimate of all users in order to compute the precoding weights. Since different user groups can have different resolutions for the channel estimates, the channel estimates for the users need to be aligned to the sub-carrier spacing of the pre-coder in order to minimize the interference between the users. Fig. 4 shows the derivation and alignment of the channel estimates of users belonging to different pre-coder groups for interference cancellation. Fig. 4 (a) and (b) show the alignment of channel estimates to derive MU pre-coders for the users belonging to a shorter sub-carrier spacing. In this case, the frequency domain channel estimates of the users with wider sub-carrier spacing are interpolated to align to the shorter sub-carrier spacing. Once aligned, the pre-coder weights for the shorter sub-carrier spacing can now be computed. Fig. 4 (c) and (d) show the alignment of channel estimates for the users belonging to a wider sub-carrier spacing. In this case, the frequency domain channel estimates of the users with shorter sub-carrier spacing are decimated to align to the wider sub-carrier spacing. Once aligned, the pre-coder weights for the wider sub-carrier spacing can now be computed. Note that this idea of channel interpolation and decimation for MU precoding is general and is applicable to any MU precoding that is based on channel estimates. In practice, resampling by a factor $r = \frac{p}{q}$, where $p$ and $q$ are integers without loss of generality, and $r > 1$ for interpolation and $r < 1$ for decimation, is achieved by first up-sampling (or interpolation) by a factor $p$ and then down-sampling (or decimation) by a factor $q$. The up-sampling can be performed using standard digital filtering techniques such as convolution with a frequency-limited impulse signal. The down-sampling is performed by selecting every $q$-th sample from the up-sampled data (or channel coefficients).

Mathematically, the proposed SLNR precoding is explained as follows. Without loss of generality, we can assume that user groups $t = 1,2,\ldots,T$ are sorted in increasing order of sub-carriers spacings. Let $h_{u,t,j}$ be the channel estimate for user $u$ in user group $t$ at sub-carrier $j$, where $u = 1,2,\ldots,K_t$, $t = 1,2,\ldots,T$, and $j = 1,2,\ldots,N_t$. Let $h_{u,t,j,r}$ denote user $u$'s channel estimates after $h_{u,t,j}$ is interpolated or decimated by a factor of $r$, where $r = \frac{p_t}{p_{t'}}$ and $t' = 1,2,\ldots,T$. Note that $t' > 1$ implies interpolation, $t' < 1$ implies decimation, and $t' = t$ implies neither interpolation nor decimation. Let us define the stacked channel matrix for user group $t$ at sub-carrier $j$ as
$$H_{t,j} = [S_t(1) \quad S_t(2) \quad \cdots \quad S_t(t) \quad \cdots \quad S_t(T-1) \quad S_t(T)],$$
where
$$S_t(t') = \left[h_{1,t',j,\frac{p_{t'}}{p_t}} \quad h_{2,t',j,\frac{p_{t'}}{p_t}} \quad \cdots \quad h_{K_{t'},t',j,\frac{p_{t'}}{p_t}}\right]. \quad (2)$$
The SLNR pre-coder for user $u$ in user group $t$ at sub-carrier $j$ is then given by
$$P_{u,t,j} = h_{u,t,j}^H (H_{t,j} H_{t,j}^H + \sigma^2 I)^{-1}, \quad (3)$$
where $\sigma^2$ is the estimated noise variance. The overall pre-coder for user group $t$ at subcarrier $j$ is then given by
$$P_{t,j} = [P_{1,t,j} \quad P_{2,t,j} \quad \cdots \quad P_{K_t,t,j}].$$

Note that in practice, the precoding is performed per subband (SB) which comprises of consecutive sub-carriers. For per SB SLNR precoding, the channel estimates in (2) and (3) should be replaced with the dominant eigenvector (associated with the largest eigenvalue) of the average covariance matrix of channel estimates within the SB.

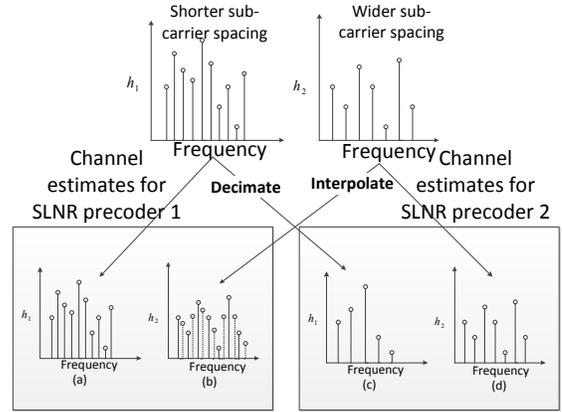

Fig. 4. Channel estimation for SLNR pre-coder design

**Complexity**

In terms of additional processing complexity to support this method, we need to add additional $M$ FFT engines at the base-station per user group, where $M$ is the number of antennas. The FFT engine size is dependent on the sub-carrier spacing. Since this complexity increase is at the base-station, additional complexity may be acceptable for an implementation. Note that the UE side, there is no additional signal processing requirement. If a UE supports multiple numerologies, it may need to dynamically configure the numerology such as FFT size, sub-carrier spacing and CP length, depending on the information provided by the eNB. Some changes to the air interface may be needed to support pilot transmissions and channel estimation when multiple numerologies are operated, which is under further investigation.

III. PERFORMANCE VALIDATION

Two users, User 1 and User 2, with different numerologies as listed in Table 1 are considered. In particular, the sub-carrier spacings for User 1 and User 2 are assumed to be 15 kHz and 30 kHz, respectively. The MU-MIMO setup for simulation is shown in Fig. 5. The dual-polarized ULA array with $0.5\lambda$ spacing in azimuth (horizontal) is considered at the eNB. The two users are assumed to be located at 135° and 45° direction, where 90° is the boresight direction to provide spatial separation. One omni-receive antenna for each user is assumed. The channel model is assumed to be 3GPP 3D channel model

[6] with 5G CDL-A power-delay profile (PDP) developed for link level performance evaluation [7]. The channel is assumed to be known at the eNB for the purpose of this analysis. This can be obtained on a per-UE basis, for example, using sounding reference signals (SRS) or channel state information – reference signals (CSI-RS) at the eNB. The simulation assumptions for the performance evaluation are summarized in Table 1.

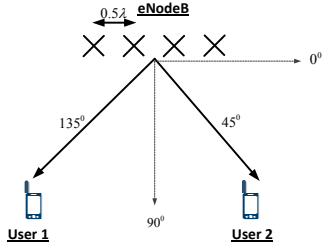

Fig. 5. MU-MIMO setup for simulation

**Table 1 Numerology and Simulation parameters**

| Parameter | User 1 | User 2 |
|---|---|---|
| Sub-carrier spacing | 15 kHz | 30 kHz |
| FFT size | 2048 | 1024 |
| Subframe duration | 1 ms | 0.5 ms |
| OFDM symbols per subframe | 14 | 14 |
| CP length (samples) | 424 | 212 |
| Channel model | 3GPP 3D channel model [6] with CDL-A PDP [7] | |
| Antenna | eNB: 2,8,16, dual-pol ULA @ 0.5 λ in azimuth<br>User 1 and 2: 1 | |
| eNB antenna polarizations | (+45°,-45°) with 0° reference in Fig. 5 | |
| Pre-coder | **MU:** CB, SLNR; **SU:** CB; Per SB precoding | |
| Channel estimation | Ideal | |
| Transmission rank | 1 | |
| Receiver | MMSE-IRC | |
| Resource block | No reference signal REs; all data REs | |

**Constellations with MU Interference**

We first investigate the impact of MU interference by observing the constellation points for the two users. Fig. 6 and Fig. 7 respectively show the constellation plots for CB and SLNR MU pre-coders for 8 and 16 antennas at the eNB. MMSE equalization and 50 dB SNR are assumed in these constellation plots. The constellation of 1 long symbol of User 1 is compared with two short symbols of User 2, which are transmitted in the same equivalent time. The two symbols of User 2 are shown separately since only the first symbol of User 2 has some overlap with the CP of the first symbol and hence, may see a different interference pattern. To quantify the MU interference, the error vector magnitude (EVM) between the received and ideal 16QAM constellations is calculated and is shown in each constellation plot. From these results, we can observe the following:

- The MU interference on long symbol (for User 1) from User 2 is greater than that on short symbol #1 (for User 2) from User 1, which in turn is similar to short symbol #2 (for User 2) from User 1. The difference may be due to the different interference pattern observed between User 1 and User 2 transmissions.
- As we increase the number of eNB antennas, the interference decreases, especially for SLNR based precoding, while the CB pre-coder is unable to take sufficient advantage of the increased antennas.
- SLNR MU pre-coder exhibits superior performance over CB MU pre-coder in terms of MU interference reduction. Also, the performance gain increases as the number of eNB antennas increases. For instance, for 16 antennas, constellation for SLNR shows almost perfect MU suppression whereas that for CB still shows large MU interference.

The results assume a fixed separation of 90° between the users for the purpose of analysis. The actual interference observed will be dependent on the separation of the users in the system, which can be accounted for by the eNB scheduler.

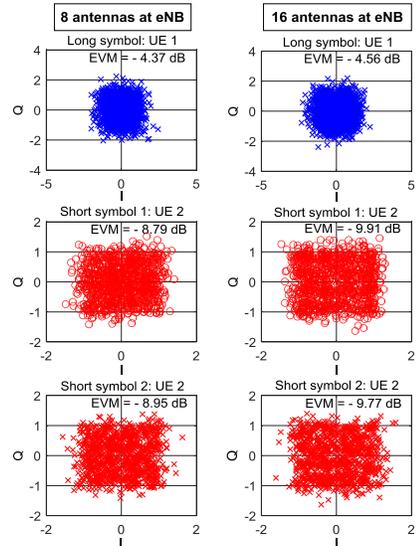

Fig. 6. Constellation points with CB MU precoding (50 dB SNR assumed)

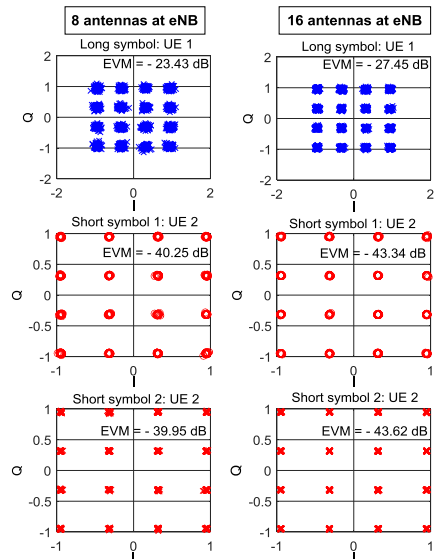

Fig. 7. Constellation points with SLNR MU precoding (50 dB SNR assumed)

## Capacity Based Simulation

Next, we compare information theoretic MU sum-capacity with the SU capacity. Dynamic switching between SU-MIMO and MU-MIMO is performed. The sum-rate capacity as a function of sub-carrier can be written as follows:

$$C(f) = \max\{C_{SU-MIMO}(f), C_{MU-MIMO}(f)\} \quad (3)$$
$$C_{SU-MIMO}(f) = \max_{k \in \{1,2\}}\{\log_2|1 + \sigma \boldsymbol{h}(f)\boldsymbol{h}_k^H(f)|\} \quad (4)$$
$$C_{MU-MIMO}(f) = \log_2(1 + SINR_1(f)) + \log_2(1 + SINR_2(f)). \quad (5)$$

Here $G_k(f)$ denotes the $1 \times N_T$ channel matrix associated with the UE-$k$ and SB $f$, $N_T$ the number of transmit antennas at the eNB, and $\sigma$ the signal power relative to noise power spectral density (SNR). Each of the SINR (signal to interference-plus-noise ratio) is computed assuming CB and SLNR MU pre-coding. The MU sum capacity is plotted as a function of SNR in Fig. 8, Fig. 9, and Fig. 10 for 2, 8, and 16 antennas at the eNB, respectively, and averaged over time and SBs which consists of four physical resource blocks (PRB).

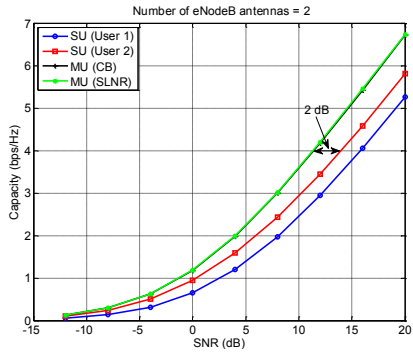

Fig. 8. MU sum-capacity for 2 antennas at eNB

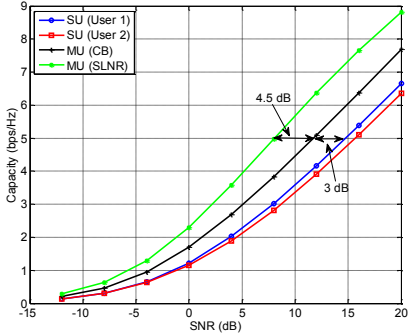

Fig. 9. MU sum-capacity for 8 antennas at eNB

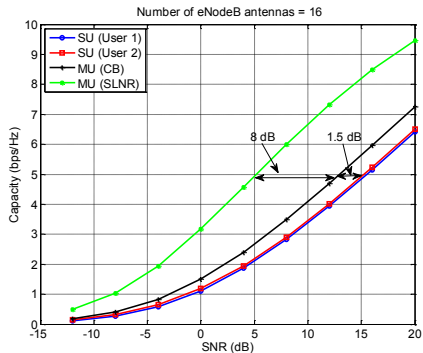

Fig. 10. MU sum-capacity for 16 antennas at eNB

*MU performance*:

Although CB precoding shows some gain, but the gain with SLNR precoding is significantly higher. For instance, for 2 antennas at the eNB, there is more than 2 dB SNR gain over SU transmission for both CB and SLNR MU precoding (Fig. 8). This is due to the well-known fact that the performance of CB and SLNR MU pre-coders are similar if the number of eNB antennas is equal to the number of MU users and each user has one receive antenna. For 8 antennas at the eNB, CB MU precoding offers 3 dB SNR gain over SU transmission whereas SLNR precoding achieves 4.5 dB additional, so 7.5 dB total SNR gain (Fig. 9). For 16 eNB antennas, however, CB MU precoding does not show similar performance gain. In fact, the gain is reduced to 1.5 dB. This is due to the fact the CB MU precoding is not designed to reduce MU interference. SLNR precoding on the other hand, maintains the performance gain and shows a total performance gain of 9.5 dB over SU (Fig.10). This is also evident from the constellations plots for the two MU pre-coders. There is no significant difference between different CB constellations (Fig. 6), but the difference is noticeable between different SLNR constellations (Fig. 7). We can thus conclude that significant performance gain can be achieved with the proposed SLNR MU precoding if the number of eNB antennas is sufficiently larger than the number of MU users.

*SU performance*:

The SU performance of User 2 is better than that of User 1 for 2 antennas at the eNB and is comparable otherwise (8 and 16 antenna cases). This is possibly due to per SB SU precoding. The subframe dAuration for User 1 is two times that for User 2 and a SB SU pre-coder is derived by considering channel covariance averaged over time (OFDM symbols) and frequency (subcarriers) within the SB.

## Link Level simulation

Finally, the link level performance evaluation is performed in order to show that the similar performance trend is maintained with actual channel coding and modulation, link-level evaluation assuming the current 3GPP LTE framework. In this evaluation, no dynamic scheduling between SU and MU transmission is assumed. This provides some insight into its performance under practical scenarios. The sum throughput vs. SNR results are shown in Fig. 11 - Fig. 13. The results verify that the performance gain of the proposed SLNR precoding is maintained in link level simulation. Note that the throughput is lower compared with capacity based evaluation due to the overhead (such as control channel, reference signal) and encoding constraints (such as discrete modulation and coding schemes) which is required for practical implementation.

The BLER vs. SNR performance for 8 antennas at eNB, MCS 9, and CDL-A channel model is also shown in Fig. 14. It can be observed that the significant MU throughput gain can be achieved over SU throughout (Fig. 12) while keeping the BLER performance within 2 dB at 10% target BLER (Fig. 14). Note that the SU-MU BLER gap can be reduced by suppressing MU interference using more number of eNB antennas.

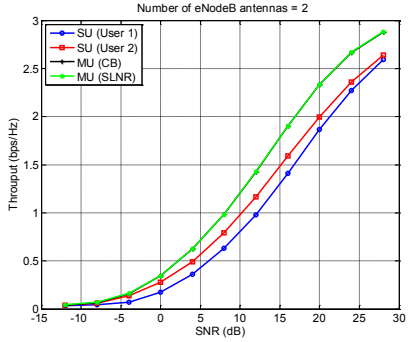

Fig. 11. Sum throughput from link level simulation for 2 antennas at eNB

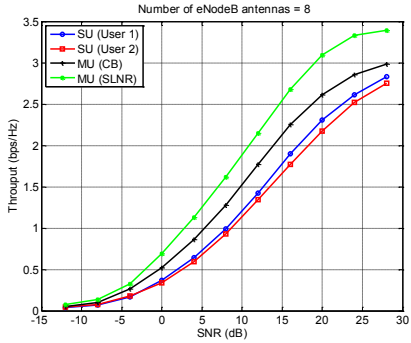

Fig. 12. Sum throughput from link level simulation for 8 antennas at eNB

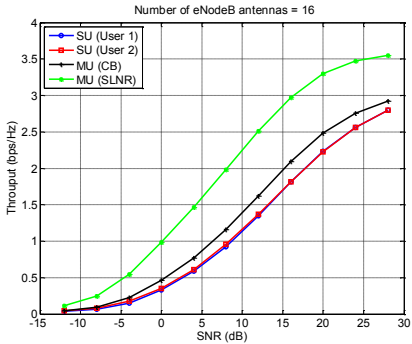

Fig. 13. Sum throughput from link level simulation for 16 antennas at eNB

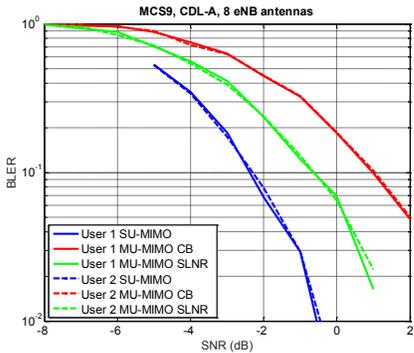

Fig. 14. BLER vs. SNR performance

## IV. CONCLUSIONS

In this paper, MU-MIMO supporting variable numerologies has been developed. We provide analysis for a two user case, in particular, with two different subcarrier spacings and with sufficient separation. A novel SLNR MU precoding method based on interpolation or decimation of interference channels is proposed to support multiple numerologies. The capacity based and link level performance evaluation is performed to demonstrate the performance gain of the proposed precoding scheme compared with the SU MIMO transmission under varying number of eNB antennas and 3GPP 3D channel model. It is shown that the proposed scheme has the potential to provide flexible numerology support with significant performance gains as the number of antennas at the eNB increases. It is worth pointing out that although simulation results assume 2-user MU-MIMO setup for simplicity, the proposed MU-MIMO transmission scheme is general and is applicable to higher-order MU-MIMO transmissions. It is expected that the performance of higher-order MU-MIMO will show similar, if not better, gains. The higher-order MU-MIMO performance evaluation can be considered in the future extension of this work.

For future studies, the availability and accuracy of channel knowledge at the transmitter needs to be verified, especially when multiple numerologies are operated in parallel. The scheduling algorithm of the eNB should also be accounted in the analysis to decide which numerologies and users are scheduled for transmission for practical deployments. Potential extensions of this work include reference signal design for channel state information (CSI) estimation and CSI feedback schemes when multiple numerologies are in effect. Another interesting future extension could be the study of the impact of different numerologies. In particular, the tradeoffs between the performance such as MU sum-rate and the extent of decimation/interpolation in the channel estimation step can be studied for different numerologies.

### ACKNOWLEDGMENT

This work was performed while the first author was with Samsung Research America. We would like to acknowledge feedback and input from Young-Han Nam and Jianzhong Zhang at Samsung Research America on the pre-coder design. We would also like to acknowledge the help from Hongbo Si for the link-level performance evaluation.